\documentstyle[epsfig,prl,aps]{revtex}
\begin{document}
\draft \twocolumn[\hsize\textwidth\columnwidth\hsize\csname
@twocolumnfalse\endcsname
\title{From vacuum nonsingular black hole to variable
cosmological constant}
\author{Irina Dymnikova
}
\address{Department of Mathematics and Computer Science,
         University of Warmia and Mazury,\\
Zolnierska 14, 10-561 Olsztyn, Poland; e-mail: irina@matman.uwm.edu.pl}

\maketitle

\begin{abstract}
In the spherically symmetric case the dominant energy condition,
together with the requirement of regularity at the center,
asymptotic flatness and finiteness of the ADM mass, defines the
family of asymptotically flat globally regular solutions to the
Einstein minimally coupled equations which includes the class of
metrics asymptotically de Sitter as $r\rightarrow 0$ and
asymptotically Schwarzschild as $r\rightarrow\infty$. A source
term connects smoothly de Sitter vacuum in the origin with the
Minkowski vacuum at infinity and corresponds to anisotropic vacuum
defined macroscopically by the algebraic structure of its
stress-energy tensor invariant under boosts in the radial
direction.
 In the range of masses $m\geq
m_{crit}$, de Sitter-Schwarzschild geometry describes  a vacuum
nonsingular black hole, and for $m<m_{crit}$ it describes G-lump
which is a vacuum self-gravitating particle-like structure without
horizons. Quantum energy spectrum of G-lump is shifted down by the
binding energy, and zero-point vacuum mode is fixed at the value
corresponding to the Hawking temperature from the de Sitter
horizon.  Space-time symmetry changes smoothly from the de Sitter
group near the center to the Lorentz group at infinity and the
standard formula for the ADM mass relates it to de Sitter vacuum
replacing a central singularity at the scale of symmetry
restoration. This class of metrics is easily extended to the case
of nonzero cosmological constant at infinity. The source term
connects then smoothly two de Sitter vacua which makes possible to
 relate a spherically symmetric anisotropic vacuum with an
$r-$dependent cosmological term $\Lambda_{\mu\nu}$.
\end{abstract}


\vskip0.2in
]

\section {Introduction}

Confrontation of models with observations in cosmology as well as
the inflationary paradigm, compellingly require treating a
cosmological constant as a variable dynamical quantity. Big value
of the cosmological constant at the very early stage of the
Universe evolution is needed to explain the reason for expansion
\cite{us75} as well as puzzles of the standard hot big bang model
\cite{olive,kolb}. The key cosmological parameter to decide if
cosmological constant is zero or not today, is the product of the
 Hubble parameter and the age of the Universe $Ht$. In the standard cosmology without
 cosmological constant this product
never exceeds the unity, but it is possible in the presence of a
nonzero cosmological constant \cite{kolb}. Therefore, if the
Hubble parameter and the age of the Universe are found in
observations to satisfy the bound $Ht> 1$, it requires a term in
the expansion rate equation that acts as a cosmological constant
\cite{kraus}. With taking into account uncertainties in models the
best fit to achieve consensus between observational constraints is
\cite{kraus,ostriker,bahcall}
$$
   H=(70-80) km~s^{-1}Mpc^{-1},~~~t=[(13-16)\pm 3] Gy,$$
   $$ \Omega_{matter}=0.3-0.4,~~~ \Omega_{\Lambda}=0.6-0.7,$$
where $\Omega=\rho_{today}/\rho_{crit}$, and the critical density
$\rho_{crit}$ corresponds to $\Omega=1$.

A cosmological term was introduced by Einstein in 1917 into his
equations describing gravity as space-time geometry (G-field)
generated by matter
    $$
    G_{\mu\nu}=-8\pi G T_{\mu\nu}\eqno(1)
    $$
to make them consistent with Mach's principle which was one of his
primary motivations \cite{eins}. Einstein's formulation of Mach's
principle was that some matter has the property of inertia only
because there exists also some other matter in the Universe
(\cite{sciama}, Ch.2). When Einstein found that Minkowski geometry
is the regular solution to (1) perfectly describing inertial
motion in the absence of any matter, he modified his equations by
adding the cosmological term $\Lambda g_{\mu\nu}$ in the hope that
modified equations
    $$
    G_{\mu\nu}+\Lambda g_{\mu\nu}=-8\pi G T_{\mu\nu}\eqno(2)
    $$
will have reasonable regular solutions only when matter is present
- if matter is the source of inertia, then in case of its absence
there should not be any inertia \cite{bondi}. The primary task of
$\Lambda$ was thus to eliminate inertia in case when matter is
absent by eliminating regular G-field solutions in case when
$T_{\mu\nu}=0$.

Soon after introducing $\Lambda g_{\mu\nu}$,  de Sitter found
quite reasonable solution with $\Lambda g_{\mu\nu}$ and without
$T_{\mu\nu}$ \cite{desitter},
      $$ ds^2=\biggl(1-\frac{\Lambda}{3}r_2\biggr)dt^2
     -\frac{dr^2}{\biggl(1-\frac{\Lambda}{3}r_2\biggr)}-r^2d\Omega^2\eqno(3)
      $$
whose nowadays triumphs are well known \cite{olive,kolb}.

 In de Sitter geometry $\Lambda$ must be constant
by virtue of the Bianchi identities $G^{\mu\nu}_{;\nu}=0$. It
plays the role of a universal repulsion whose physical sense
remained obscure during several decades when de Sitter metric has
been mainly used in quantum field theory as a simple testing
ground for developing the quantum field technics in curved
space-time.

Almost fifty years later, in 1965, two papers appeared in the same
issue of the Soviet Physics JETP, which shed some light on the
physical nature of the de Sitter geometry. The first was the paper
by Sakharov \cite{sakharov}, in which he suggested that
gravitational effects dominate the equation of state of a cold
baryon-lepton superdense matter at  densities $\rho\sim{10^{74} g~
cm^{-3}}$ (GUT density is of order  $\rho\sim{10^{77} g~ cm^{-3}}$
for GUT scale $\sim{10^{15}}$GeV) and that one of possible
equations of state in such a regime could be
      $$
        p=-\rho\eqno(4)
      $$
formally corresponding to the equation of state for $\Lambda
g_{\mu\nu}$ shifted to the right hand side of the Einstein
equation (2) as some stress-energy tensor. The physical sense of
this operation has been clarified in the second 1965 paper by
Gliner \cite{gliner} who interpreted $\Lambda g_{\mu\nu}$ as
corresponding to a stress-energy tensor of a superdense vacuum
      $$
        T_{vac}^{\mu\nu}=(8\pi G)^{-1}\Lambda g_{\mu\nu}\eqno(5)
      $$
In the Petrov classification scheme \cite{petrov} stress-energy
tensors are classified on the basis of their algebraic structure.
When the elementary divisors of the matrix $T_{\mu\nu}-\lambda
g_{\mu\nu}$ are real, the eigenvectors of $T_{\mu\nu}$ are
non-isotropic and form a comoving reference frame with the
timelike vector representing a velocity. A comoving reference
frame is defined uniquely if and only if none of the spacelike
eigenvectors $\lambda_a (a=1,2,3)$ coincides with a timelike
eigenvalue $\lambda_0$. A stress-energy tensor (5) with all
eigenvalues equal, has an infinite set of comoving reference
frames and hence no preferred one. An observer moving through de
Sitter vacuum (5) cannot in principle measure his velocity with
respect to it, since an observer's comoving frame is also comoving
for (5) \cite{gliner}. Gliner suggested that at superhigh
densities a continual medium is formed with attraction between its
elements, which is phenomenologically described by stress-energy
tensor (5) with the negative pressure and the equation of state
(4). The other very important hypothesis suggested by Gliner in
this paper was that such a state could be achieved in a
gravitational collapse \cite{gliner}.

In 1967 De Witt found that quantum effects in one-loop
approximation lead to the vacuum stress-energy tensor of a form
\cite{dewitt}
       $$<0|T_{\mu\nu}|0>=\rho_{vac} g_{\mu\nu}\eqno(6)
       $$
In 1968 Zel'dovich proposed to relate $\Lambda g_{\mu\nu}$ to
gravitational interaction of virtual particles in vacuum
\cite{zeld}.

In  80-s several attempts have been made to eliminate a black hole
singularity by replacing it at the Planck scale curvature with de
Sitter metric using direct matching of Schwarzschild metric
outside to de Sitter metric inside a short spacelike transitional
layer of the Planckian depth
\cite{markov,bernstein,farhi,shen,valera}. The matched solutions
typically have a jump at the junction surface which comes from
singularity of a tangential pressure there. The situation with de
Sitter-Schwarzschild transition has been analyzed by Poisson and
Israel  who suggested to introduce a transitional layer of
"noninflationary material" of uncertain depth in which geometry
can be self-regulatory and describable semiclassically down to a
few Planckian radii by the Einstein equations with a source term
representing vacuum polarization effects \cite{werner}.

Generic properties of "noninflationary material" have been
considered in Ref.\cite{me92} for the case of a smooth transition
from de Sitter vacuum at the origin to Minkowski vacuum at
infinity, and the exact analytical solution has been found
describing a vacuum nonsingular black hole in a simple
semiclassical model for vacuum polarization in the gravitational
field. In the course of Hawking evaporation such a black hole
evolves towards a self-gravitating particle-like vacuum structure
without horizons \cite{me96}, kind of gravitational vacuum soliton
called G-lump \cite{mass}.

Model-independent analysis of the Einstein spherically symmetric
minimally coupled equations has shown \cite{mass} which kind of
geometry they can describe in principle (no matter which matter
source is responsible for a stress-energy tensor) if certain
general requirements are satisfied:

a) regularity of metric and density at the center;

b) asymptotic flatness at infinity and finiteness of the ADM mass;

c) dominant energy condition for $T_{\mu\nu}$.

The requirements (a)-(c) define the family of asymptotically flat
solutions with the regular center which includes the class of
metrics asymptotically de Sitter as $r\rightarrow 0$ and
asymptotically Schwarzschild as $r\rightarrow\infty$. A source
term belongs to the class  of stress-energy tensors invariant
under boosts in the radial direction and connects de Sitter vacuum
in the origin with the Minkowskli vacuum at infinity. Space-time
symmetry changes smoothly from de Sitter group at the center to
the Lorentz group at infinity through the radial boosts in
between, and the standard formula for the ADM mass relates it
(generically, since a matter source can be any from considered
class) to both de Sitter vacuum replacing a singularity  and
breaking of space-time symmetry.

 This class of metrics can be extended to the case of non-zero
cosmological term at infinity \cite{us97} corresponding to
extension of the Einstein cosmological term $\Lambda g_{\mu\nu}$
to an $r-$dependent second rank symmetric tensor
$\Lambda_{\mu\nu}$\cite{me00} connecting in a smooth way two de
Sitter vacua with different values of a cosmological constant.

This talk is organized as follows. In Section II we outline de
Sitter-Schwarzschild geometry, and in Section III its extension to
the case of non-zero cosmological constant at infinity. In Section
IV we present variable cosmological term $\Lambda_{\mu\nu}$ and
quantum energy spectrum of G-lump.
 In Section V we outline the results concerning connection between the
ADM mass and cosmological term. Section VI contains discussion.

\section {De Sitter-Schwarzschild geometry}

A static spherically symmetric line element can be written in
the standard form (see, e.g., \cite{tolman}, p.239)
     $$
     ds^2 = e^{\mu(r)}dt^2 - e^{\nu(r)} dr^2 - r^2 d\Omega^2\eqno(7)
     $$
where $d\Omega^2$ is the metric of a unit 2-sphere.

The Einstein equations (1) reduce
to (\cite{tolman}, p.244)
    $$
    8\pi G T_t^t = 8\pi G\rho(r)= e^{-\nu}\biggl(\frac{{\nu}^{\prime}}{r}
    -\frac{1}{r^2}\biggr)
    +\frac{1}{r^2}\eqno(8)
    $$
    $$8\pi G T_r^r =-8\pi G p_r(r)= -e^{-\nu} \biggl(\frac{{\mu}^{\prime}}{r}
     +\frac{1}{r^2}\biggr)
    +\frac{1}{r^2}\eqno(9)
    $$
    $$8\pi G T_{\theta}^{\theta}=8\pi G T_{\phi}^{\phi}=-8\pi G p_{\perp}(r)=$$
    $$-e^{-\nu}\biggl(\frac{{{\mu}^{\prime\prime}}}{2}
    +\frac{{{\mu}^{\prime}}^2}{4}
    +\frac{({{\mu}^{\prime}-{\nu}^{\prime}})}{2r}-\frac{{\mu}^{\prime}
    {\nu}^{\prime}}{4}\biggr)\eqno(10)
    $$
Here $\rho(r)=T^t_t$ is the energy density (we adopted $c=1$ for
simplicity), $p_r(r)=-T^r_r$ is the radial pressure, and
$p_{\perp}(r)=-T_{\theta}^{\theta}=-T_{\phi}^{\phi}$ is the
tangential pressure for anisotropic perfect fluid (\cite{tolman},
p.243). A prime denotes differentiation  with respect to $r$.

Integration of Eq.(8) gives \cite{wald}
    $$e^{-\nu(r)}=1-\frac{2GM(r)}{r};~~M(r)
    =4\pi\int_0^r{\rho(x)x^2dx}\eqno(11)
    $$
which has for large $r$ the Schwarzschild asymptotic
$e^{-\nu}=1-{2Gm}/{r}$, where the mass parameter $m$ is given by
   $$
   m=4\pi\int_0^{\infty}{\rho(r) r^2 dr}\eqno(12)
   $$

Analysis of this system in the case when requirements (a)-(c) are
satisfied leads to the following results \cite{mass}:

The dominant energy condition $T^{00}\geq|T^{ab}|$ for each $a,b=1,2,3$,
which holds if and only if \cite{HE}
     $$\rho\geq0;~~~~-\rho\leq p_k\leq \rho;~~~~k=1,2,3\eqno(13)
     $$
implies that the local energy density is non-negative and each
principal pressure never exceeds the energy density. In the limit
$r\rightarrow\infty$ the condition of finiteness of the mass (12)
requires density profile $\rho(r)$ to vanish at infinity quicker
than $r^{-3}$, and the dominant energy condition (13) requires
both radial and tangential pressures to vanish as
$r\rightarrow\infty$. Then $\mu^{\prime}=0$ and $\mu=$const at
infinity, and the standard boundary condition $\mu\rightarrow 0$
as $r\rightarrow \infty$ leads to asymptotic flatness needed to
identify (12) as the ADM mass \cite{wald}. As a result we get the
Schwarzschild asymptotic at infinity
     $$T_{\mu\nu}=0;~~ds^2=\biggl(1-\frac{2Gm}{r}\biggr)-
     \frac{dr^2}{\biggl(1-\frac{2Gm}{r}\biggr)}-r^2d\Omega^2\eqno(14)
     $$
From Eq.(8)-(10) we derive the equation (see also \cite{apj})
     $$
     p_{\perp}=p_r+\frac{r}{2}p_r^{\prime}+(\rho+p_r)\frac{G M(r)
     +4\pi G r^3 p_r}{2(r-2G M(r))}\eqno(15)
     $$
which is generalization of the Tolman-Oppenheimer-Volkoff
equation (\cite{wald}, p.127)
to the case of different principal pressures, and
the equation \cite{oppi}
     $$ T_t^t-T_r^r=p_r+\rho=\frac{1}{8\pi G}\frac{e^{-\nu}}{r}
     (\nu^{\prime}+\mu^{\prime})\eqno(16)
     $$
From Eq.(11) it follows that for any regular value of $e^{\nu(r)}$
$M(r)=0$ at $r=0$ and thus $\nu(r)\rightarrow 0$ as $r\rightarrow
0 $ \cite{oppi}. The dominant energy condition allows us to fix
asymptotic behavior of a mass function and of a metric at
approaching the regular center. Requirement of regularity of
density $\rho(r=0)<\infty$, leads, by Eq. (13), to regularity of
pressures. Requirement of regularity of the metric,
$e^{\nu(r)}<\infty$, leads then, by (16), to
$\nu^{\prime}+\mu^{\prime}=0$ and $\nu+\mu=\mu(0)$ at $r=0$ with
$\mu(0)$ playing the role of the family parameter.

The weak energy condition,
$T_{\mu\nu}\xi^{\mu}\xi^{\nu}\geq 0$
for any timelike vector $\xi^{\mu}$, which
is satisfied
if and only if
$\rho\geq 0; \rho + p_k \geq 0, k=1,2,3$
and which is contained in the dominant energy condition\cite{HE},
defines, by Eq.(16), the sign of the
sum $\mu^{\prime}+\nu^{\prime}$. In the case when $e^{\nu (r)}>0$
everywhere, it demands  $\mu^{\prime}+\nu^{\prime}\geq 0$ everywhere.
In case when $e^{\nu (r)}$ changes sign, the function
$T_t^t-T_r^r$ is zero, by Eq.(16), at the horizons where $e^{-\nu}=0$.
In the regions inside the horizons, the radial coordinate $r$
is timelike and $T_t^t$ represents a tension,
$p_r=-T_t^t$, along the axes of the spacelike 3-cylinders of constant time
$r$=const \cite{werner}, then $T_t^t-T_r^r=-(p_r+\rho)$,
and the weak energy condition  still
demands $\nu^{\prime}+\mu^{\prime} \geq 0$ there.
As a result the function $\mu+\nu$ is a function growing from $\mu=\mu(0)$
at $r=0$ to $\mu=0$ at $r\rightarrow\infty$, which gives $\mu(0)\leq 0$.

The well known example of solution from this family is boson stars
\cite{boson} (for review \cite{Mielke}).

The range of family parameter $\mu(0)$ dictated by the weak energy
condition, includes the value $\mu(0)=0$, which corresponds to
$\nu+\mu=0$ at the center. In this case the function
$\phi(r)=\nu(r)+\mu(r)$ is zero at $r=0$ and at
$r\rightarrow\infty$, its derivative is non-negative, it follows
that $\phi(r)=0$, i.e., $\nu(r)=-\mu(r)$ everywhere. The weak
energy condition defines also equation of state and thus
asymptotic behavior as $r\rightarrow 0$. The function
$\phi(r)=\mu(r)+\nu(r)$, which is equal zero everywhere for $0\leq
r<\infty$, cannot have extremum at $r=0$, therefore
$\mu^{\prime\prime}+\nu^{\prime\prime}=0$ at $r=0$ (this is easily
proved by contradiction using the Maclaurin rule for even
derivatives in the extremum). It leads, by using L'Hopital rule in
Eq.(16), to $p_r+\rho=0$ at $r=0$. In the limit $r\rightarrow 0$
Eq.(15) becomes $p_{\perp}=-\rho-\frac{r}{2}\rho^{\prime}$. The
energy dominant condition (13) requires $\rho^{\prime}\leq 0$,
while regularity of $\rho$ requires $p_k+\rho<\infty$ and thus
$|\rho^{\prime}|<\infty$. Then the equation of state near the
center becomes $p=-\rho$, which gives de Sitter asymptotic (3) as
$r\rightarrow 0$ \cite{mass}.

Summarizing, we conclude that if we require asymptotic flatness,
regularity of a density and metric  at the center and finiteness
of the ADM mass, then the dominant energy condition defines the
family of asymptotically flat solutions with the regular center
which includes the class of metrics
    $$e^{\mu (r)}=e^{-\nu (r)}=g(r)=1-2G M(r)/r \eqno(19)$$
with $M(r)$ given by Eq.(11), whose behavior in the origin -
asymptotically de Sitter as $r\rightarrow 0$, is dictated by  the
weak energy condition.

For this class a source term has the algebraic structure
$$
T_t^t=T_r^r:~~~T_{\theta}^{\theta}=T_{\phi}^{\phi} \eqno(20)
$$
and the equation of state is
     $$p_r=-\rho:~~~p_{\perp}=-\rho-(r/2)\rho^{\prime}\eqno(21)
     $$
It connects de Sitter vacuum $T_{\mu\nu}=\rho_0 g_{\mu\nu}$ in the
origin with the Minkowski vacuum $T_{\mu\nu}=0$ at infinity, and
generates de Sitter-Schwarzschild geometry \cite{me96}
asymptotically de Sitter as $r\rightarrow 0$ and asymptotically
Schwarzschild as $r\rightarrow\infty$.

 Note, that
if we postulate regularity also for pressures, then the weak
energy condition is enough to distinguish the class of metrics
(19) \cite{mass}.

The weak energy condition $p_{\perp}+\rho\geq 0$ gives
$\rho^{\prime}\leq 0$, so that it demands monotonic decreasing of
a density profile. By Eq.(10) it leads to the important fact that,
except the point $r=0$ where $g(r)$ has the maximum, in any other
extremum $g^{\prime\prime}> 0$, so that the function $g(r)$ has in
the region $0<r<\infty$ only minimum and the metric (19) can have
not more than two horizons \cite{mass}.

To find explicit form of $M(r)$ we have to choose some density
profile leading to the needed behavior of $M(r)$ as $r\rightarrow
0$, $M(r)\simeq{(4\pi/3) \rho_0 r^3}$. The simplest choice
\cite{me92}
       $$\rho(r)=\rho_0 e^{-r^3/r_0^2 r_g};
       ~~~r_0^2=3/\Lambda;~~~r_g=2Gm\eqno(22)
       $$
can be interpreted \cite{me96} as due to vacuum polarization in
the spherically symmetric gravitational field as described
semiclassically by the Schwinger formula $w\sim{e^{-F_{crit}/F}}$
(see, e.g., \cite{igor}) with tidal forces $F\sim{r_g/r^3}$ and
$F_{crit}\sim{1/r_0^2}$, in agreement with the basic idea
suggested by Poisson and Israel that in Schwarzschild-de Sitter
transition space-time geometry can be self-regulatory as a result
of  vacuum polarization effects \cite{werner}.

 The key point is the existence
of two horizons, a black hole event horizon $r_{+}$ and an
internal horizon $r_{-}$. A critical value of a mass parameter
exists, $m_{crit}$, at which the horizons come together and which
puts a lower limit on a black hole mass \cite{me96}.
 For the model (22)
      $$
       m_{crit}\simeq{0.3 m_{Pl}\sqrt{\rho_{Pl}/\rho_0}}\eqno(23)
      $$
De Sitter-Schwarzschild configurations are shown in Fig.1.
\begin{figure}
\vspace{-8.0mm}
\begin{center}
\epsfig{file=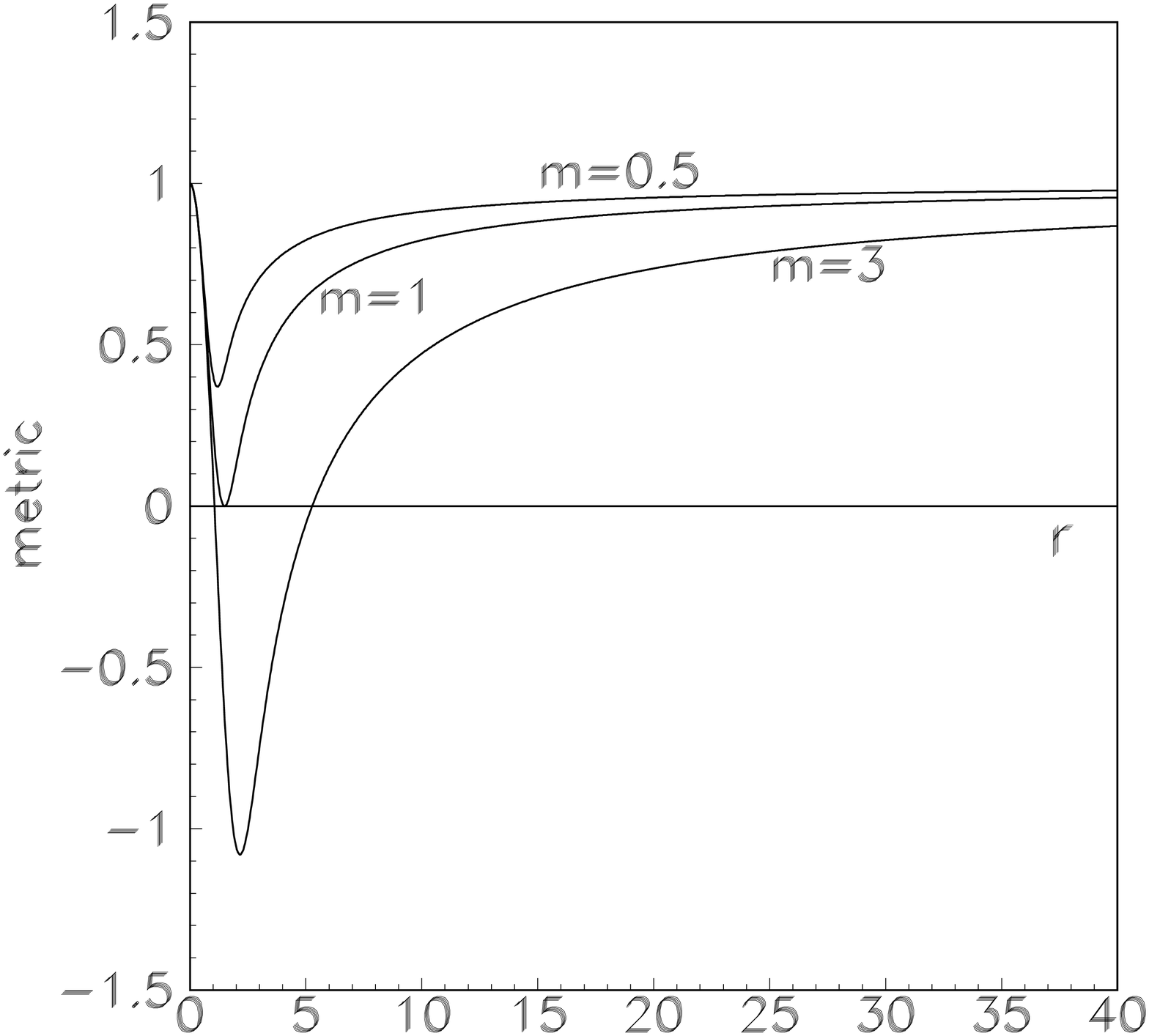,width=8.0cm,height=5.5cm}
\end{center}
\caption{ The metric $g(r)$ for de Sitter-Schwarzschild
configurations plotted for the case of the density profile (22).
The mass parameter $m$ is normalized to $m_{crit}$. }
\label{fig.1}
\end{figure}

 For $m\geq m_{crit}$ de
Sitter-Schwarzschild geometry describes the vacuum nonsingular
black hole ($\Lambda$BH) \cite{me92}, and global structure of
space-time, shown in Fig.2 \cite{me96}, contains an infinite
sequence of black and white holes whose future and past
singularities are replaced with regular cores $\cal{RC}$
asymptotically de Sitter as $r\rightarrow 0$.

\begin{figure}
\vspace{-8.0mm}
\begin{center}
\epsfig{file=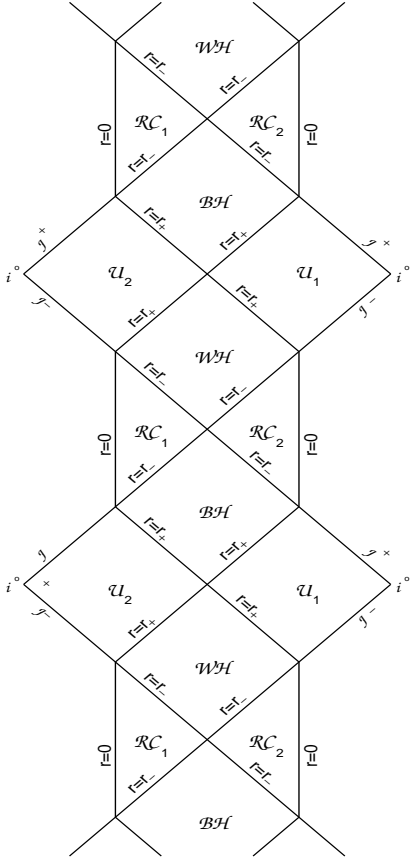,width=8.0cm,height=12.9cm}
\end{center}
\caption{Penrose-Carter diagram for $\Lambda$ black hole.}
\label{fig.2}
\end{figure}

A $\Lambda$BH emits Hawking radiation from both black hole and
cosmological horizons with the Gibbons-Hawking temperature
$T=\hbar \kappa(2\pi kc)^{-1}$\cite{GH} where $\kappa$ is the surface
gravity and $k$ is the Boltzmann constant. For a $\Lambda$BH the
temperature from horizons is given by \cite{me96}
           $$
              T_h=\frac{\hbar G} {2\pi c k r_0}\biggl(\frac{M(r_h)}{r_h^2}
              -\frac{M^{\prime}(r_h)}{r_h}\biggr)\eqno(24)
            $$
In the limit $r_g/r_0\gg 1$, the temperature tends to the
Schwarzschild value $T_{Schw}=\hbar c^3/8\pi G k m$ on the black
hole horizon and to the de Sitter value $T_{DeS}=-\hbar c/2\pi k
r_0$ on the internal horizon.While a $\Lambda$BH loses its mass,
horizons come together, and configuration evolves towards a
self-gravitating particle-like structure without horizons
\cite{me96}.
 Temperature-mass diagram is shown in Fig.3.
Its form is generic for de Sitter-Schwarzschild geometry and does
not depend on particular form of a density profile. The
temperature $T_{+}$ on BH horizon $r_{+}$ is positive by general
laws of BH thermodynamics \cite{wald}. It drops to zero at
$m=m_{crit}$, while the Schwarzschild asymptotic requires
$T_{+}\rightarrow 0$ as $m\rightarrow\infty$. As a result the
temperature-mass diagram should have a maximum between $m_{crit}$
and $m\rightarrow\infty$ \cite{me96}. In a maximum a specific heat
is broken and changes sign testifying to a second-order phase
transition in the course of Hawking evaporation and suggesting
symmetry restoration to the de Sitter group in the origin
\cite{me97}.

\begin{figure}
\vspace{-8.0mm}
\begin{center}
\epsfig{file=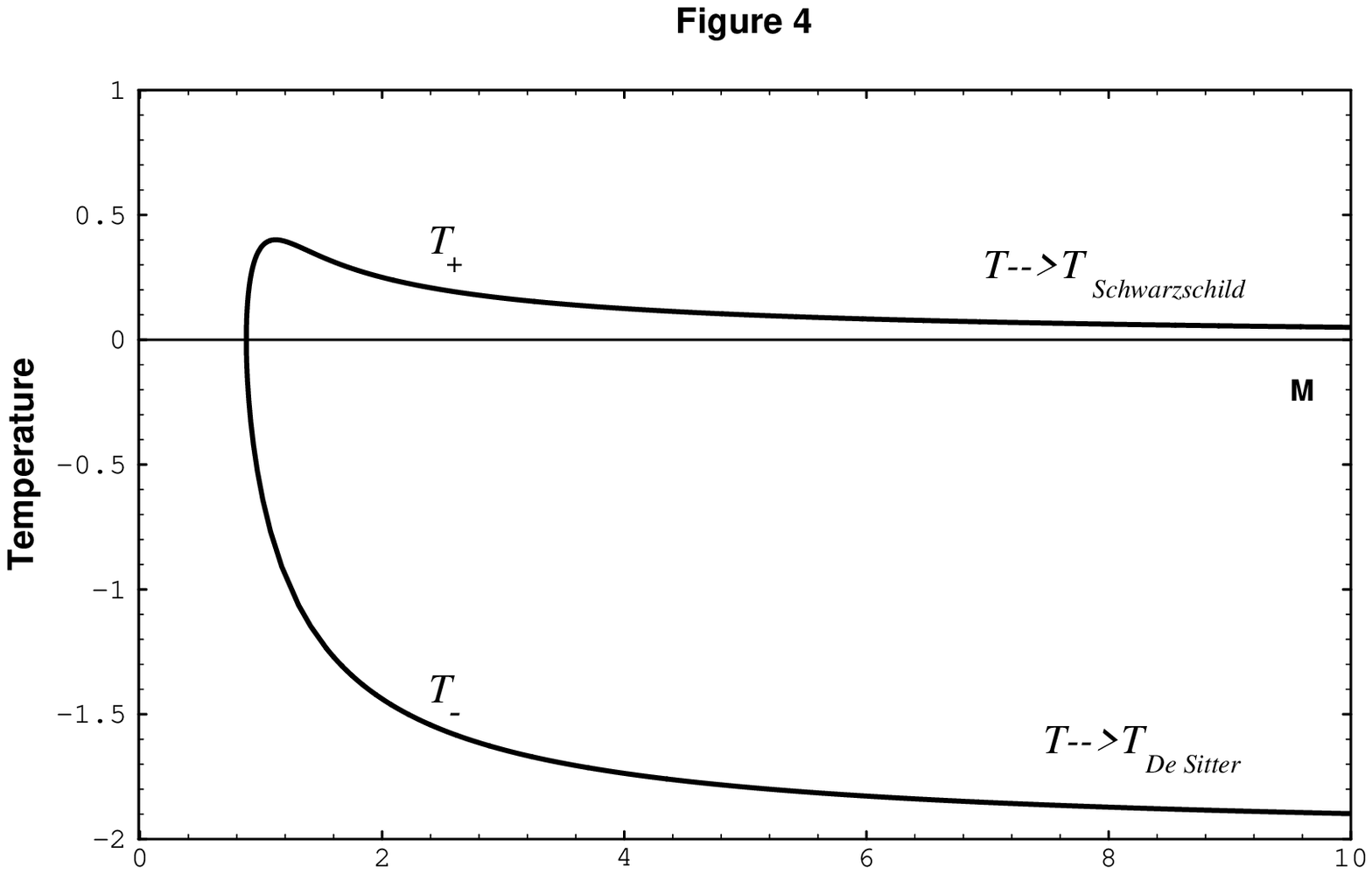,width=8.0cm,height=5.5cm,clip=}
\end{center}
\caption{
Temperature-mass diagram for $\Lambda$ black hole.}
\label{fig.3}
\end{figure}

For masses $m<m_{crit}$ de Sitter-Schwarzschild geometry describes
a self-gravitating particle-like vacuum structure without
horizons, globally regular and globally neutral.
 It resembles
Coleman's lumps - non-singular, non-dissipative solutions of
finite energy, holding themselves together by their own
self-interaction \cite{lump}. The lump idea goes back to the
Einstein proposal to describe an elementary particle by regular
solution of nonlinear field equations as "bunched field" located
in the confined region where field tension and energy are
particularly high \cite{bunch}. De Sitter-Schwarzschild lump is
regular solution to the Einstein equations, perfectly localized
(see Fig.4) in a region where field tension and energy are
particularly high (this is the region of the former singularity),
so we can call it G-lump \cite{mass}.
\begin{figure}
\vspace{-8.0mm}
\begin{center}
\epsfig{file=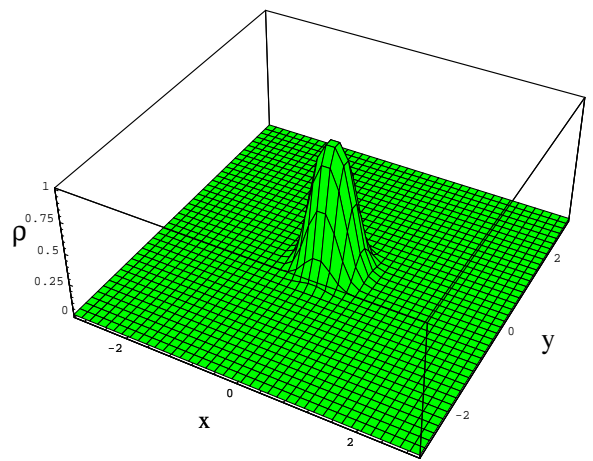,width=8.0cm,height=6.5cm}
\end{center}
\caption{
G-lump in the case $r_g=0.1r_0$ ($m\simeq{0.06 m_{crit}})$.
}
\label{fig.4}
\end{figure}

 It holds itself together by gravity
due to balance between gravitational attraction outside and
gravitational repulsion inside of zero-gravity surface $r=r_c$
beyond which the strong energy condition of singularities theorems
\cite{HE}, $(T_{\mu\nu}-T g_{\mu\nu}/2)\xi^{\mu}\xi^{\nu})\geq 0$,
is violated \cite{me96}. The surface of zero gravity is defined by
$2\rho+r\rho^{\prime}=0$. It is depicted in Fig.5 together with
horizons and with the surface $r=r_s$ of zero scalar curvature
$R(r_s)=0$ which represents the characteristic curvature size in
the de Sitter-Schwarzschild geometry. In the case of the density
profile (22) the characteristic size $r_s$ is given by
      $$r_s=\biggl(\frac{4}{3}r_0^2 r_g\biggr)^{1/3}
     =\biggl(\frac{m}{\pi\rho_0}\biggr)^{1/3}\eqno(25)
     $$
and confines about 3/4 of the mass $m$.
\begin{figure}
\vspace{-8.0mm}
\begin{center}
\epsfig{file=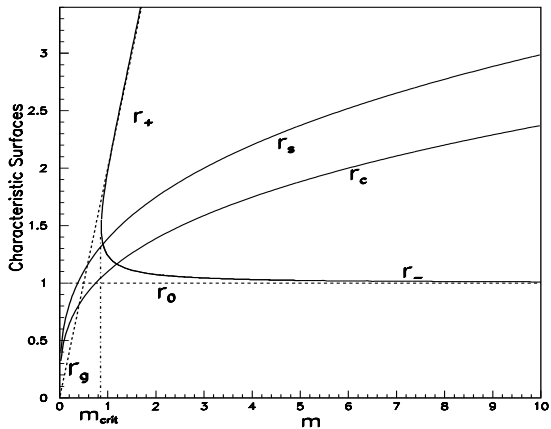,width=8.0cm,height=6.5cm}
\end{center}
\caption{
Horizons of $\Lambda$BH, surface of zero scalar curvature $r=r_s$
and surface of
zero gravity $r=r_c$.
}
\label{fig.5}
\end{figure}

The question of stability of de Sitter-Schwarzschild
configurations is currently under investigation. De
Sitter-Schwarzschild black hole configuration obtained by direct
matching of the Schwarzschild metric outside to de Sitter metric
inside of a spacelike three-cylindrical short transitional layer
\cite{valera} is a stable configuration in a sense that the
three-cylinder does not tend to shrink down under perturbations
\cite{Poisson}. De Sitter-Schwarzschild configurations considered
above represent general case of a smooth transition with a
distributed density profile. The heuristic argument in favor of
their stability comes from comparison of the ADM mass with the
proper mass \cite{wald} $\mu=4\pi\int_0^{\infty}{\rho(r)\biggl(1
-\frac{2 G M(r)}{r}\biggr)^{-1/2}r^2 dr}$. In the spherically
symmetric case the ADM mass represents the total energy,
$m=\mu+$binding energy \cite{wald}. In de Sitter-Schwarzschild
geometry $\mu$ is bigger than $m$. This suggests that the
configuration might be stable since energy is needed to break it
up \cite{me00}. Analysis of stability of a $\Lambda$BH as an
isolated system by Poincare's method, with the total energy $m$ as
a thermodynamical variable and the inverse temperature as the
conjugate variable \cite{Kaburaki}, shows immediately its
stability with respect to spherically symmetric perturbations. The
analysis by Chandrasekhar method \cite{chandra} is straightforward
for a $\Lambda$BH stability to external perturbations, in close
similarity with the Schwarzschild and Reissner-Nordstr\"om cases.
The potential barriers in one-dimensional wave equations governing
perturbations, external to the event horizon, are real and
positive, and stability follows from this fact \cite{chandra}.
Preliminary results suggest stability also for the case of G-lump.
In the context of catastrophe-theory analysis, de
Sitter-Schwarzschild configuration resembles high-entropy neutral
type in the Maeda classification, in which a non-Abelian structure
may be approximated as a sphere of uniform vacuum density
$\rho_{vac}$ whose radius is the Compton wavelength of a massive
non-Abelian field, and self-gravitating particle approaches the
particle solution in the Minkowski space \cite{Maeda}.

\section{Two-Lambda geometry}

The class of metrics (19)-(20 is easily extended to the case of
nonzero background cosmological constant $\lambda$, by introducing
               $$T_t^t(r)=\rho(r)+(8\pi G)^{-1}\lambda \eqno(26)
               $$
Then the metric function $g(r)$ in Eq.(19) is given by \cite{us97}
        $$g(r)=1-\frac{2GM(r)}{r} -\frac{\lambda r^2}{3}\eqno(27)
        $$
For $r\ll (3r_g/\Lambda)^{1/3}$, the metric (27) behaves like de
Sitter metric with cosmological constant $\Lambda+\lambda$, while
for $r\gg (3r_g/\Lambda)^{1/3}$ it aproaches the Kottler-Trefftz
metric \cite{kot}
     $$
       ds^2=\biggl(1-\frac{r_{g}}{r}-\frac{\lambda r^2}{3}\biggr)dt^2
        -\biggl(1-\frac{r_{g}}{r}-\frac{\lambda r^2}{3}\biggr)^{-1}dr^2$$
          $$-r^2(d\vartheta^2+\sin^2\vartheta d\varphi^2),\eqno(28)$$
which is frequently referred to in the literature as the
Schwarzschild-de Sitter geometry describing cosmological black
hole. The metric (27) represents thus its nonsingular
modification.

The two-lambda space-time has in general three horizons: a
cosmological horizon $r_{++}$, a black hole horizon $r_{+}$ and an
internal horizon $r_{-}$ which  can be formally identified as the
Cauchy horizon (see also \cite{werner}) as formed by zero
generators inextendible to the past.  The metric function (27) is
plotted in Fig.6.
\begin{figure}
\vspace{-8.0mm}
\begin{center}
\epsfig{file=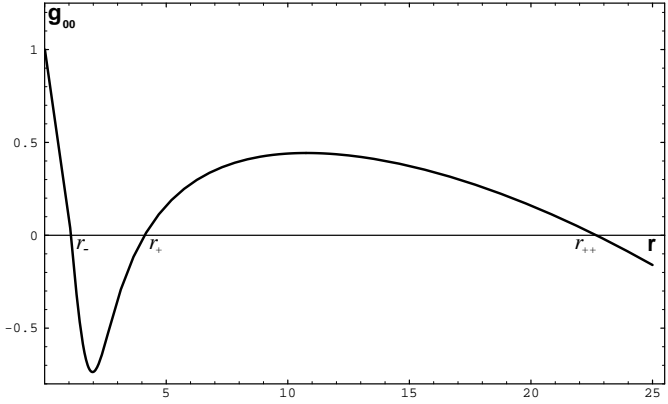,width=9.0cm,height=5.5cm}
\end{center}
\caption{
 Two-lambda spherically symmetric solution for the case
 $~ q=\sqrt{\Lambda/\lambda}=10~ $ and $~m=4\sqrt{{3}/{\Lambda}}
 ({c^2}/{G})$.
}
\label{fig.6}
\end{figure}
In the range of horizons $r_h\ll({\Lambda r_g}/{3})^{1/3}$ the
internal horizon is given approximately by \cite{us97}
      $$
       r_{-}\simeq{\sqrt{\frac{3}{\Lambda+\lambda}}}
          \biggl[1+\frac{1}{4 r_g}\sqrt{\frac{3}{\Lambda+\lambda}}
           \biggl(\frac{\Lambda}{\Lambda+\lambda}\biggr)^2$$
        $$\biggl[1+\frac{5}{4 r_g}\sqrt{\frac{3}{\Lambda+\lambda}}
   \biggl(\frac{\Lambda}{\Lambda+\lambda}\biggr)^2\biggr]\biggr]\eqno(29)
   $$
for $r_g\gg{\sqrt{{3}/{\Lambda+\lambda}}
({\Lambda}/{\Lambda+\lambda})^2}$.

In the range of $r_h\gg({\Lambda r_g}/{3})^{1/3}$, the
cosmological horizon is located approximately at
    $$
    r_{++}\simeq{\sqrt{\frac{3}{\lambda}}-\frac{r_g}{2}}\eqno(30)
    $$
for $r_g\ll\sqrt{{3}/{\lambda}}({\Lambda}/{\lambda})$.

In the interface a horizon can be written in the form
$r_h=r_g+\varepsilon, ~\varepsilon\ll r_g$ which gives
the black hole horizon
      $$
       r_{+}\simeq {r_g\biggl[1+\frac{\lambda r_g^2}{3}
      -\exp{\biggl(-\frac{\Lambda r_g^2}{3}\biggr)}\biggr]}\eqno(31)
      $$
for $r_g$ within the range $\sqrt{{3}/{\Lambda}}\ll r_g \ll
\sqrt{{3}/{\lambda}}$.

Horizon-mass diagram is plotted in Fig.7 for the case of the
density profile given by (22).
\begin{figure}
\vspace{-8.0mm}
\begin{center}
\epsfig{file=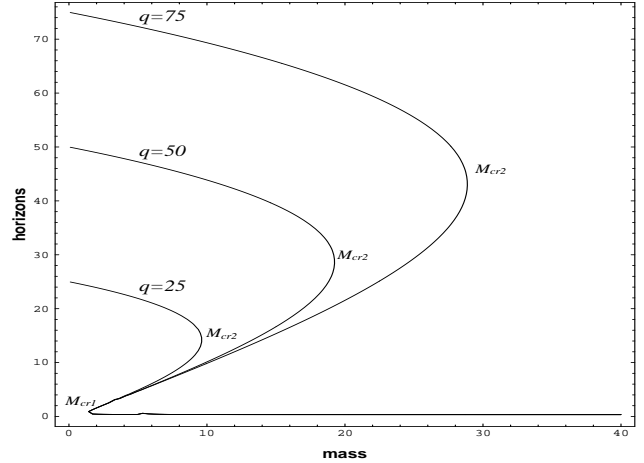,width=9.0cm,height=6.5cm}
\end{center}
\caption{ Horizon-mass diagram for the metric (27). The parameter
$M$ is the mass $m$ normalized to $(3/G^2\Lambda)^{1/2}$.
 }
\label{fig.7}
\end{figure}
There are two critical values of the mass $m$, restricting the
mass of a nonsingular cosmological black hole from below and from
above. Within the range of masses $m_{cr1}<m<m_{cr2}$, the metric
(27) has three horizons and describes a nonsingular cosmological
black hole. Its global structure is shown in Fig.8.
\begin{figure}
\vspace{-8.0mm}
\begin{center}
\epsfig{file=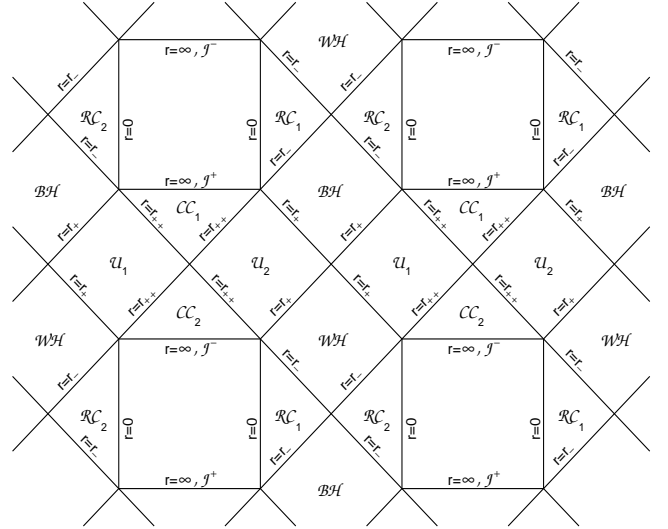,width=8.9cm,height=7.3cm}
\end{center}
\caption{ Penrose-Carter diagram for two-lambda black hole. }
 \label{fig.8}
\end{figure}
This diagram is similar to the case of Reissner-Nordstr\"om-de
Sitter geometry \cite{bron}. The essential difference is that the
timelike surface $r=0$ is regular in our case. The global
structure of nonsingular cosmological black hole contains an
infinite sequence of asymptotically de Sitter (small background
$\lambda$) universes $~{\cal U}_{1}$, $~{\cal U}_{2}$, black and
white holes $~{\cal BH}$, $~{\cal WH}~$ whose singularities are
replaced with future and past regular cores $~{\cal RC}_1$, ${\cal
RC}_2~$ (with $~\Lambda~+\lambda~$ at $r\rightarrow 0$), and
"cosmological cores" $~{\cal CC}~$ (regions between cosmological
horizons and spacelike infinities). Rectangular regions confined
by the surfaces $r=0$  and $r=\infty$ do not belong to the
diagram.

Specification of these regions \cite{us98} is given by
 the invariant quantity \cite{igor,victor}
$\Delta=g^{\mu\nu}r,_{\mu}r,_{\nu}$.
Dependently on the sign of $\Delta$, space-time is divided into
$~R~$ and $~T~$ regions (see \cite{igor,victor}):
In the $~ R~$ regions the normal vector to the surfaces $r=$const,
$N_{\mu}=r,_{\mu}$ is spacelike, and an observer on those surfaces
can send radial signals directed to both inside and outside of them.
In the $~T~$ regions the normal vector $N_{\mu}$ is timelike,
surfaces $r=$const are spacelike, and both signals propagate on the
same side of this surface, and any observer can cross
the surface $r=$const only once and only in the same direction.
 The $~R~$ and $~T~$ regions are separated by horizons, where
$\Delta=0$. For the two-lambda space-times, the regions $~{\cal
RC}$ and $~\cal U$ are $~R~$ regions, while  the regions
$~\cal{BH}$, $~\cal{WH}$ and $~\cal {CC}$ are $~T~$ regions. For
the metric in the Kruskal form $~\Delta=(1/2)g_{00}^{-1}r,_u
r,_v~$; in the $~ T~$ regions $~\Delta>0~$, the vector $r,_u$
cannot be zero, and the conditions $r,_u>0$ and $r,_u<0$ are
invariant \cite{victor}. For $r,_u<0$, $~T~$ region is $~T_{-}~$
region of contraction, so that $~\cal{BH}$ are $T_{-}$ regions,
while $ ~\cal{WH}$ are expanding $T_{+}$ regions, since $r,_u>0$
there.

 Five types of globally regular spherically symmetric configurations
described by two-lambda geometry,
are plotted in Fig.9 for the case
$q\equiv{\sqrt{{\Lambda}/{\lambda}}}=10$.
\begin{figure}
\vspace{-8.0mm}
\begin{center}
\epsfig{file=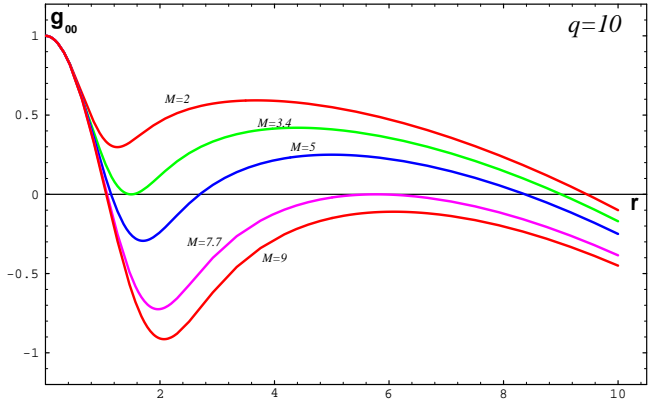,width=9.0cm,height=6.0cm}
\end{center}
\caption{ Two-lambda configurations for the case $q=10$. The
parameter $M$ is a mass $m$ normalized to $(3/G^2 \Lambda
)^{1/2}$.
}
\label{fig.9}
\end{figure}

 The critical value of mass
at which the internal horizon $r_{-}$ coincides with the black
hole horizon $r_{+}$, defines the first extreme black hole state.
The value $m_{cr1}$ puts the lower limit for a black hole mass. It
practically does not depend on the parameter
$q=\sqrt{\Lambda/\lambda}$ and is given by (23). This extreme
black hole is shown in Fig.9 (the curve $M=3.4$). The
Schwarzschild-de Sitter family of singular black holes contains
masses between zero and the size of the cosmological horizon (see,
e.g.,\cite{bousso}). Replacing a black hole singularity with a
cosmological constant $\Lambda$ results in appearance of the lower
limit on a mass of cosmological black hole which is almost the
same as in the case of nonsingular $\Lambda$ black hole at the
Minkowski space background \cite{me96}. It represents the new type
spherically symmetric configuration - the extreme neutral
nonsingular cosmological black hole whose internal horizon
coincides with a black hole horizon.

An upper limit $m_{cr2}$, at which  the black hole horizon $r_{+}$
coincides with the cosmological horizon $r_{++}$, corresponds to
the nonsingular modification of the Nariai solution \cite{nariai},
with the additional internal horizon which is absent in the Nariai
case. The value of $~m_{cr2}~$ depends essentially on the
parameter $q=\sqrt{\Lambda/\lambda}$ (see Fig.7).

Beyond the limiting masses $m_{cr1}$ and $m_{cr2}$, there exist
two different types of globally regular spherically symmetric
configurations:

(i) A spherically symmetric self-gravitating particle-like
structure at the de Sitter background in the range of masses
$m<m_{cr1}$. This G-lump differs from the case of Minkowski space
background Fig.1 by existence of the cosmological horizon.

(ii) The case $m>m_{cr2}$ differs essentially from the
Schwarzschild-de Sitter case by existence of an internal horizon.
Configuration of this type which we called "de Sitter bag"
\cite{us97}, corresponds to cosmology with the same global
structure as for de Sitter geometry, but with cosmological
constant smoothly evolving from $\Lambda$ in the past to $\lambda$
in the future.

\section{Variable cosmological term}

Stress-energy tensors (20) for the considered class of metrics
belong to the Petrov type [(II)(II)].
 The first symbol in the brackets denotes the eigenvalue
related to the timelike eigenvector representing a velocity.
Parentheses combine equal eigenvalues. A stress-energy tensor of
this type  has an infinite set of comoving reference frames, since
it is invariant under boosts in the radial direction, and can be
thus identified as describing a spherically symmetric anisotropic
vacuum (an observer moving through such  a medium cannot in
principle measure the radial component of his velocity with
respect to it), i.e., vacuum
 with variable energy density and pressures, macroscopically
defined by the algebraic structure  of its stress-energy tensor
$T_{\mu\nu}^{vac}$ \cite{me92}.
 In the case of nonzero background $\lambda$
it connects smoothly  two de Sitter vacua with different values of
cosmological constant. This makes it possible to interpret
$T_{\mu\nu}^{vac}$ as corresponding to the extension of the
algebraic structure of the cosmological term from $\Lambda
g_{\mu\nu}$ (with $\Lambda$=const) to an $r$-dependent
cosmological term $\Lambda_{\mu\nu}=8\pi G T_{\mu\nu}^{vac}$,
evolving from $\Lambda_{\mu\nu}=\Lambda g_{\mu\nu}$ as
$r\rightarrow 0$ to $\Lambda_{\mu\nu}=\lambda g_{\mu\nu}$ as
$r\rightarrow\infty$, and satisfying the equation of state (21)
with $8\pi G \rho^{\Lambda}=\Lambda^t_t$, $8\pi G
p_r^{\Lambda}=-\Lambda^r_r$ and $8\pi G p_{\perp}^{\Lambda}
=-\Lambda^{\theta}_{\theta}=-\Lambda^{\phi}_{\phi}$ \cite{me00}.

In quantum field theory cosmological constant $\Lambda$ is related
to zero-point vacuum energy. A zero-point energy of G-lump which
clearly represents an elementary spherically symmetric excitation
of a vacuum defined macroscopically by (20), can be evaluated in
simple quantum minisuperspace model \cite{mass}. Since de Sitter
vacuum is trapped within a G-lump, we can model it by a spherical
bubble whose density decreases with a distance.  In the
Finkelstein coordinates, de Sitter-Schwarzschild geometry is
described by the metric
    $$
     ds^2=d\tau^2-\frac{2GM(r(R,\tau))}{r(R,\tau)}-r^2(R,\tau)d\Omega^2
     \eqno(32)
    $$
The equation of motion
$\dot{r}^2 + 2r\ddot{r}-8\pi G\rho(r)r^2=f(R)$ \cite{us00},
where dot denotes differentiation with respect to $\tau$ and
$f(R)$ is constant of integration, has the first integral
         $$
          \dot{r}^2 - \frac{2GM(r)}{r}=f(R)\eqno(33)
         $$
which resembles the equation of a particle in the potential
$V(r)=-\frac{GM(r)}{r}$, with the constant of integration $f(R)$
playing the role of the total energy $f=2E$.

A spherical bubble can be described by the minisuperspace model
with a single degree of freedom \cite{vil}. The momentum operator
is introduced by $\hat{p}=-i{l_{Pl}}^2 d/dr$, and the equation
(33) transforms into the Wheeler-DeWitt equation in the
minisuperspace \cite{vil} which reduces to the Schr\"odinger
equation
        $$
        \frac{\hbar^2}{2m_{Pl}}\frac{d^2\psi}{dr^2}-(V(r)-E)\psi=0\eqno(34)
        $$
with the potential (in the Planckian units)
        $$
         V(r)=-\frac{GM(r)}{r}\eqno(35)
         $$
depicted in Fig.10.
\begin{figure}
\vspace{-8.0mm}
\begin{center}
\epsfig{file=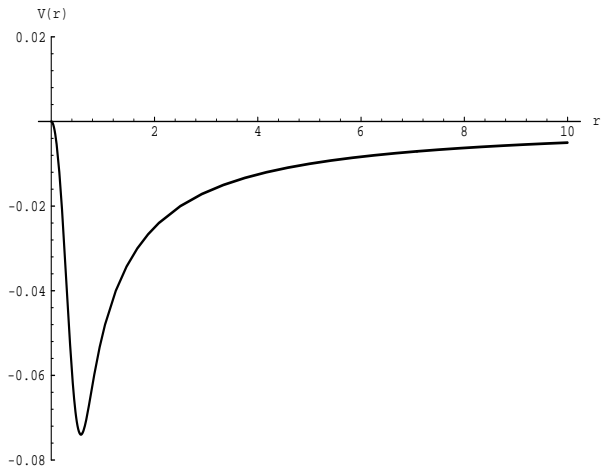,width=8.0cm,height=6.5cm}
\end{center}
\caption{The plot of the potential (35) for G-lump with $r_g=0.1
r_0 (m\simeq 0.07 m_{crit})$.
}
\label{fig.10}
\end{figure}

Near the minimum $r=r_m$ the potential takes the form
$V(r)=V(r_m)+4\pi G p_{\perp}(r_m)(r-r_m)^2$. Introducing the
variable $x=r-r_m$ we reduce  Eq.(34) to the equation for a
harmonic oscillator
           $$
           \frac{d^2\psi}{dx^2}-\frac{m_{Pl}^2\omega^2 x^2}{\hbar^2}\psi
            +\frac{2m_{Pl}\tilde{E}}{\hbar^2}\psi=0\eqno(36)
           $$
where $\tilde{E}=E-V(r_m)$, $\omega^2=\Lambda c^2
\tilde{p}_{\perp}(r_m)$, and $\tilde{p}_{\perp}$ is the
dimensionless pressure normalized to vacuum density $\rho_0$ at
$r=0$; for the density profile (22)
$\tilde{p}_{\perp}(r_m)\simeq{0.2}$. The energy spectrum
               $$
               E_n=\hbar \omega \biggl(n+\frac{1}{2}\biggr)
                -\frac{GM(r_m)}{r_m}E_{Pl}\eqno(37)
                $$
is shifted down by the minimum of the potential $V(r_m)$ which
represents the binding energy. The energy of zero-point vacuum
mode \cite{mass}
            $$
             \tilde{E}_0=\frac{\sqrt{3\tilde{p}_{\perp}}}{2}
             \frac{\hbar c}{r_0}\eqno(38)
            $$
never exceeds the binding energy $V(r_m)$. It
remarkably agrees with the Hawking temperature from the de Sitter horizon
$kT_H=\frac{1}{2\pi}\frac{\hbar c}{r_0}$ \cite{GH}, representing the energy of
virtual particles which could become real in the presence of the horizon.
In the case of G-lump which is structure without horizons, kind of
gravitational vacuum exciton,
they are confined by the binding energy $V(r_m)$.

\section {Cosmological term as a source of mass}

The mass of both G-lump and $\Lambda$BH is directly connected to
cosmological term $\Lambda_{\mu\nu}$ by the ADM formula (12) which
in this case reads
           $$
             m=(2 G)^{-1} \int_0^{\infty}{\Lambda_t^t(r) r^2 dr}\eqno(39)
           $$
and relates mass to the de Sitter vacuum at the origin (which is
thus evidently trapped within an object) \cite{mass}.

The Minkowski geometry allows existence of inertial mass as the
Lorentz invariant $m^2=p_{\mu}p^{\mu}$  of a  test body. High
symmetry of this geometry allows both existence of inertial frames
and of quantity $m$ as the measure of inertia, but geometry tells
nothing about this quantity.

In the Schwarzschild geometry the parameter $m$ is responsible for
geometry, it is identified as a gravitational mass of a source by
asymptotic behavior of the metric at infinity. By the equivalence
principle, gravitational mass is equal to inertial mass. The
inertial mass is represented thus by a purely geometrical
quantity, the Schwarzschild radius $r_g$ which  is geometrical
fact of the Schwarzschild geometry \cite{wheeler}. However it
still does not tell about origin of a mass.

 The geometrical fact of de Sitter-Schwarzschild geometry is that
a mass $m$ (identified by Schwarzschild asymptotic at infinity) is
related to cosmological term, since Schwarzschild singularity is
replaced with a de Sitter vacuum. The operation of introducing
mass by the ADM formula (39) is impossible in the de Sitter
geometry, since symmetry of the source term $T_{\mu\nu}=\rho_0
g_{\mu\nu}=(8\pi G)^{-1}\Lambda g_{\mu\nu}$ is too high and
$\rho_0$=const everywhere.  In the case of de Sitter-Schwarzschild
geometry symmetry of a source term is reduced from the full
Lorentz group to the Lorentz boosts in the radial direction only.
Together with asymptotic flatness this allows introducing a
distinguished point as the center of an object whose ADM mass is
defined by the standard formula (12). The reduced symmetry of a
source and the asymptotic flatness of geometry are responsible for
mass of an object given by (39).

Let us note that this observation does not depend on
identification of a vacuum tensor of the algebraic structure (20)
as corresponding to  variable cosmological term
$\Lambda_{\mu\nu}$. Any stress-energy tensor for this class of
metrics (no matter interpreted as $\Lambda_{\mu\nu}$ or not) is
invariant under full Lorentz group in the origin and at infinity
but only under radial boosts in between. And for any source from
this class the standard formula (12) for the ADM mass relates it
to both de Sitter vacuum trapped in the origin and  breaking of
space-time symmetry.

 This picture seems to be in remarkable
conformity with the basic idea of the Higgs mechanism for
generation of mass via spontaneous breaking of symmetry of a
scalar field vacuum from a false vacuum (where
$T_{\mu\nu}=V(0)g_{\mu\nu}$, and $p=-\rho$), to a true vacuum
$T_{\mu\nu}=0$. In both cases de Sitter vacuum is involved and
vacuum symmetry is broken. Even graphically the gravitational
potential $g(r)$ resembles a Higgs potential (see Fig.11).
\begin{figure}
\vspace{-8.0mm}
\begin{center}
\epsfig{file=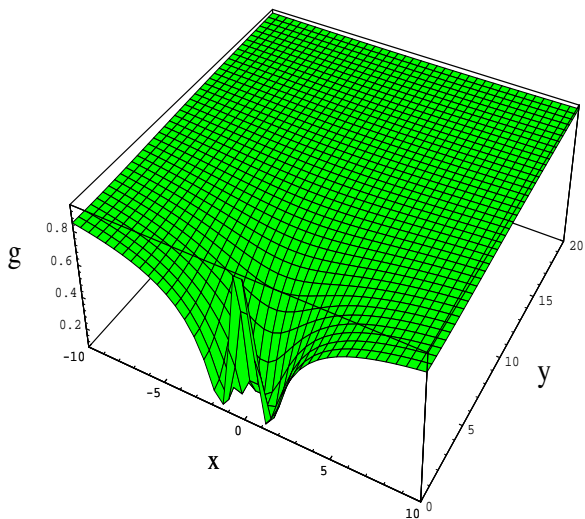,width=8.0cm,height=7.5cm}
\end{center}
\caption{The gravitational potential $g(r)$ for the case of
G-lump with the mass a little bit less than $m_{crit}$.
}
\label{fig.11}
\end{figure}

The difference is that in case of a mass coming from
$\Lambda_{\mu\nu}$ by (39), the gravitational potential $g(r)$ is
generic, and de Sitter vacuum supplies a particle with a mass via
smooth breaking of space-time symmetry from the de Sitter group in
its center to the Lorentz group at its infinity.

This leads to the natural assumption \cite{ethz} that whatever
would be particular mechanism for mass generation, a fundamental
particle (a particle which does not display substructure, like a
lepton or quark) may have an internal vacuum core (at the scale
where it gets mass) related to its mass and a  geometrical size
defined by gravity. Such a core with de Sitter vacuum at the
origin and Minkowski vacuum at infinity can be approximated by de
Sitter-Schwarzschild geometry. Characteristic size in this
geometry is given by (25). It depends on vacuum density at $r=0$
and presents modification of the Schwarzschild radius $r_g$ to the
case when singularity is replaced by de Sitter vacuum. While
application of the Schwarzschild radius to elementary particle
size is highly speculative since obtained estimates are many
orders of magnitude less than $l_{Pl}$, the characteristic size
$r_s$ gives reasonable numbers (e.g., $r_s\sim{10^{-18}}$ cm for
the electron getting its mass from the vacuum at the electroweak
scale) close to estimates obtained in experiments (see
Fig.12\cite{ethz} where they are compared with electromagnetic
(EM) and electroweak (EW) experimental limits).
\begin{figure}
\vspace{-8.0mm}
\begin{center}
\epsfig{file=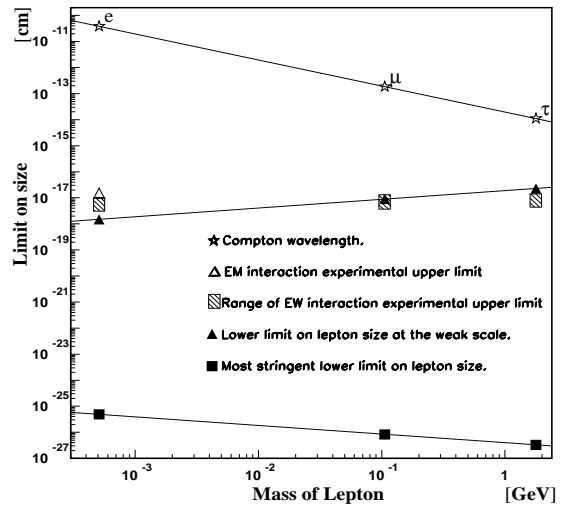,width=8.0cm,height=7.5cm}
\end{center}
\caption{Characteristic sizes for leptons \protect\cite{ethz}.
}
\label{fig.12}
\end{figure}

\section{Discussion}

The main point outlined here is the existence of the class of
globally regular solutions to the minimally coupled GR equations
(8)-(10), with a source term of the algebraic structure (20)
interpreted as spherically symmetric anisotropic vacuum with
variable density and pressures $T_{\mu\nu}^{vac}$ associated with
a time-dependent and spatially inhomogeneous cosmological term
$\Lambda_{\mu\nu} =8\pi GT_{\mu\nu}^{vac}$, whose asymptotic in
the origin, dictated by the weak energy condition, is the Einstein
cosmological term $\Lambda g_{\mu\nu}$.

 The key difference of $\Lambda_{\mu\nu}$
from the quintessence which is introduced as a time-varying
spatially inhomogeneous component of matter content with negative
pressure is in the algebraic structure of stress tensors.
Quintessence is defined by the equation of state $p=-\alpha\rho$
with $\alpha<1$ \cite{quint}. This corresponds to such a
stress-energy tensor $T_{\mu\nu}$ for which a comoving reference
frame is defined uniquely. The quintessence represents thus a
non-vacuum negative-pressure isotropic alternative to a
cosmological constant $\Lambda$ while the cosmological tensor
$\Lambda_{\mu\nu}$ represents the extension of the algebraic
structure of the Einstein cosmological term $\Lambda g_{\mu\nu}$
which makes it variable and anisotropic.

 De Sitter-Schwarzschild
geometry  (19) describes generic properties
 of any configuration satisfying (20) and requirements (a)-(c), obligatory
 for any particular model in the same sense as de Sitter geometry (3) is
 obligatory for any matter source satisfying (4).

In the inflationary cosmology which is based on generic properties
of de Sitter vacuum $\Lambda g_{\mu\nu}$ independently on where
$\Lambda$ comes from \cite{us75}, several mechanisms are
investigated relating  $\Lambda g_{\mu\nu}$ to matter sources (for
review see \cite{olive,andrej}). Most frequently considered is a
scalar field
       $$
         S=\int{d^4 x \sqrt{-g}\biggl[R
           +(\partial \phi)^2-2V(\phi)\biggr]}\eqno(40)
       $$
where $R$ is the scalar curvature,
$(\partial \phi)^2=g^{\mu\nu}{\partial }_{\mu}\phi {\partial }_{\nu}\phi$,
with various forms for a scalar field potential  $V(\phi)$.

The question whether a regular black hole  can be obtained as a
false vacuum configuration described by (40), has been addressed
in the paper  \cite{dima}, where "the no-go theorem" has been
proved: Asymptotically flat regular black hole solutions are
absent  in the theory  (40) with any non-negative potential
$V(\phi)$. This result has been extended to the case of any
$V(\phi)$ and any asymptotic and then generalized to the case of a
theory with the action $S=\int{d^4 x\sqrt{-g}\biggl[R+F[(\partial
\phi)^2,\phi]\biggr]}$, where $F$ is an arbitrary function
\cite{kirill}, to the multi-scalar theories of sigma-model type,
and to scalar-tensor and curvature-nonlinear gravity theories
\cite{kirill2}. It has been shown that the only possible regular
solutions are either de Sitter-like with a single cosmological
horizon  or those without horizons, including asymptotically flat
ones. The latter do not exist for $V(\phi)\geq 0$, so that the set
of causal false vacuum structures is the same as known for
$\phi=const$ case, namely Minkowski (or anti-de Sitter),
Schwarzschild, de Sitter, and Schwarzschild-de Sitter
\cite{kirill,kirill2}, and thus does not include de
Sitter-Schwarzschild configurations.

In the case of {\it complex} massive scalar field the regular
structures can be obtained in the minimally coupled theory with
positive $V(\phi)$ \cite{Schunck}. These are boson stars
(\cite{Mielke} and references therein), but in this case algebraic
structure of the stress-energy tensor \cite{Mielke} does not
satisfy Eq.(20), and asymptotic  at $r=0$ is not de Sitter.

The considered connection between r-dependent cosmological term
$\Lambda_{\mu\nu}$ and the ADM mass seems to satisfy Einstein's
version of Mach's principle - no matter, no inertia - if we
explicitly separate two aspects of the problem of inertia:
existence of inertial frames and existence of inertial mass. In
empty space, $T_{\mu\nu}^{vac}=0$, inertial frames exist due to
high symmetry of Minkowski geometry, but to prove it we need a
measure of inertia, a test particle with the inertial mass, i.e. a
region in space where Minkowski vacuum is a little bit disturbed.
For the considered class of metrics with the regular center
$T_{\mu\nu}^{vac}\neq 0$ and the inertial mass is generically
related to both reduced symmetry of a source term (20) (no matter
interpreted as $\Lambda_{\mu\nu}$ or not) and de Sitter vacuum
trapped in the origin. In other words, full symmetry of Minkowski
space-time is responsible for existence of inertial frames, while
its breaking to Lorentz boosts in the radial direction only is
responsible for inertial mass.

\vskip0.1in
{\bf Acknowledgement}

 This work was supported by the Polish Committee for Scientific
Research through the Grant 5P03D.007.20, and through the grant for
UWM.


\end{document}